\documentclass[final]{aipproc}
\layoutstyle{8x11double}
\usepackage{amssymb}
\begin{document}
\title{Local-duality QCD sum rule for the pion elastic form factor}
\author{Irina Balakireva}{address={D.~V.~Skobeltsyn Institute of Nuclear
Physics, Moscow State University, 119991, Moscow, Russia}}
\author{Wolfgang Lucha}{address={Institute for High Energy Physics,
Austrian Academy of Sciences, Nikolsdorfergasse 18, A-1050 Vienna, Austria}}
\author{Dmitri Melikhov}{address={Institute for High Energy Physics,
Austrian Academy of Sciences, Nikolsdorfergasse 18, A-1050 Vienna,
Austria},altaddress={Faculty of Physics, University of Vienna,
Boltzmanngasse 5, A-1090 Vienna, Austria}}\maketitle

The goal of this work was to study the accuracy of the pion form
factor derived from the so-called local-duality (LD) version of
QCD sum rules \cite{radyushkin}: (i) It is based on a dispersive
three-point sum rule at $\tau\,=\,0$ (i.e., infinitely large Borel
mass parameter); then all power corrections vanish and the form
factor is just given by a perturbative spectral representation
with a cut applied at an effective threshold $s_{\rm eff}(Q)$.
(ii) It makes use of a model for $s_{\rm eff}(Q)$ based on some
smooth interpolation between its values at $Q\,\to \,0$ determined
by the Ward identity and at $Q\,\to \,\infty$ determined by
factorization. For instance, in \cite{braguta}, a simple
interpolation formula, $s_{\rm
eff}(Q)=4\pi^2f^2_{\pi}/[1+\alpha_s(Q)/\pi],$ has been proposed.

Obviously, the LD model is an approximate model that does not take
into account the details of the confinement dynamics, and it is
important to understand its accuracy. The only property of the
theory relevant for this model is the factorization of hard form
factors. Consequently, the model may be tested in quantum
mechanics for potentials containing both Coulomb and confining
interactions. The corresponding analysis has been reported in
\cite{irina}; the main conclusions obtained are:\vspace{.5ex}

\noindent 1. For $Q^2\lesssim1\;\mbox{GeV}^2$, the exact effective
threshold found from comparison with the data does depend on $Q^2$
\cite{lms_sr1,lms_sr2}, and it exhibits a rapid variation with
$Q^2$:\vspace{-.4ex}

\centerline{\includegraphics[width=6cm]{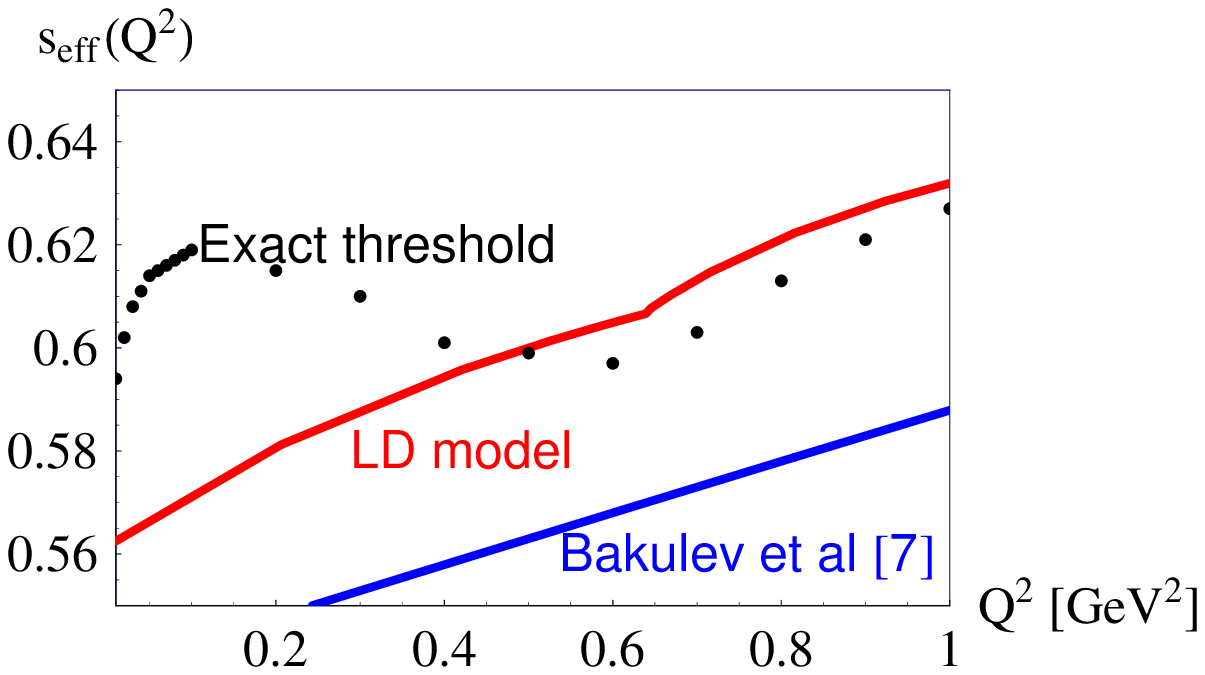}}

\noindent The accuracy of the LD model for the pion form factor~in
this region is not overwhelmingly high.\vspace{.5ex}

\noindent 2. For $Q^2\gtrsim4-6\;\mbox{GeV}^2$, the LD model is
expected to provide a good description of the pion form factor,
with an accuracy better than 20\% \cite{irina}. Moreover, the
accuracy increases rather fast with $Q^2$. Our predictions for
$F_\pi(Q^2)$ from the LD model are shown in the plot below:

\centerline{\includegraphics[width=6cm]{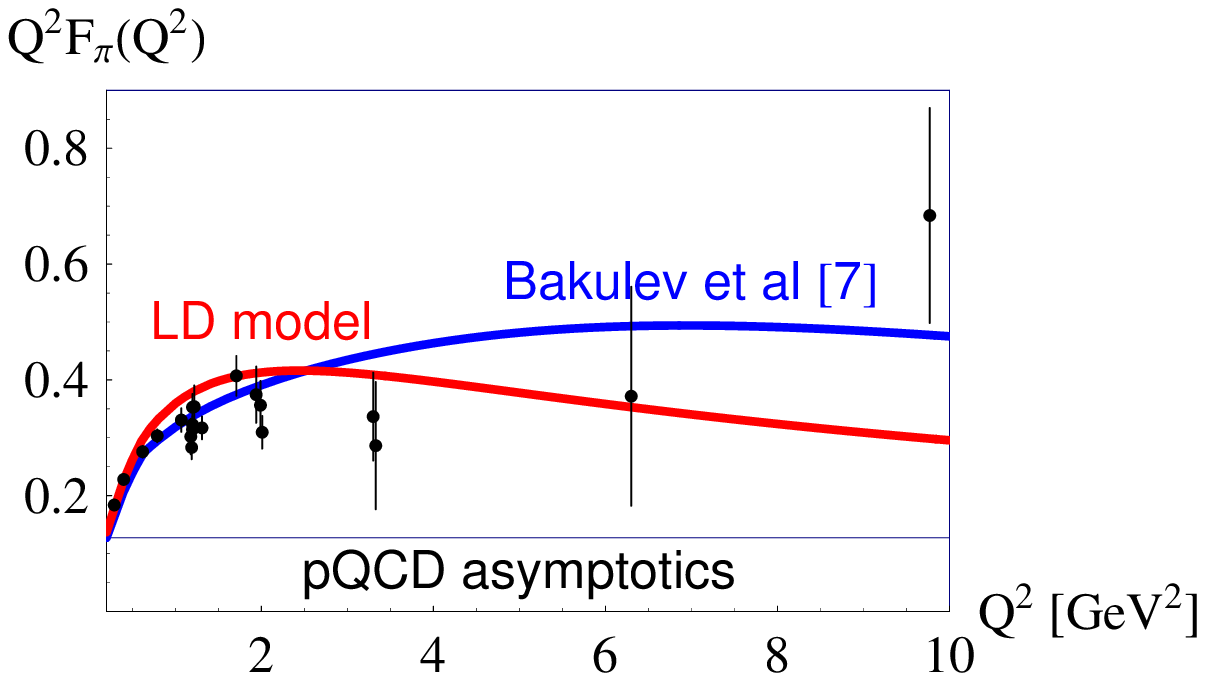}}

\noindent Our findings agree well with results from the dispersion
approach \cite{anis} but are considerably lower than the results
of sum rules with non-local condensates \cite{bakulev}.
Presumably, the discrepancy might be traced back to the procedure
of fixing the $\tau$-independent effective threshold in
\cite{bakulev} based on merely the stability criterion. As
demonstrated in \cite{lms_sr1}, stability is not sufficient to
guarantee a reliable extraction of hadron parameters.

\vspace{.23cm}\noindent{\it Acknowledgments.} D.~M.\ acknowledges
support by the Austrian Science Fund (FWF) under Project
no.~P20573.

\vspace{-.25cm}
\bibliographystyle{aipproc}

\begin{thebibliography}{30}
\vspace{-.2cm}
\bibitem{radyushkin}
A.~V.~Radyushkin, Acta Phys.~Polon.~{\bf B26}, 2067 (1995).
\bibitem{braguta}
V.~Braguta, W.~Lucha, and D.~Melikhov, Phys.~Lett.~{\bf B661}, 354
(2008); W.~Lucha and D.~Melikhov, arXiv:0812.0323.
\bibitem{irina}I.~Balakireva, arXiv:1009.4140.
\bibitem{lms_sr1}
W.~Lucha, D.~Melikhov, and S.~Simula, Phys.~Rev.~{\bf D76}, 036002
(2007); Phys.~Lett.~{\bf B657}, 148 (2007); Phys.~Atom.\
Nucl.~{\bf 71}, 1461 (2008); Phys.~Lett.~{\bf B671}, 445 (2009);
D.~Melikhov, Phys.~Lett.~{\bf B671}, 450 (2009); W.~Lucha and
D.~Melikhov, Phys.~Rev.~{\bf D73}, 054009 (2006); Phys.\
Atom.~Nucl.~{\bf 70}, 891 (2007).
\bibitem{lms_sr2}W.~Lucha, D.~Melikhov, H.~Sazdjian, and S.~Simula,
Phys.~Rev.~{\bf D80}, 114028 (2009); W.~Lucha, D.~Melikhov, and
S.~Simula, Phys.~Rev.~{\bf D79}, 096011 (2009); J.~Phys.\ {\bf
G37}, 035003 (2010); Phys.~Atom.~Nucl.~{\bf 73}, 1770 (2010);
arXiv:1003.1463; Phys.~Lett.~{\bf B687}, 48 (2010);
arXiv:1008.2698.
\bibitem{anis}
V.~V.~Anisovich, D.~I.~Melikhov, and V.~A.~Nikonov,
Phys.~Rev.~{\bf D52}, 5295 (1995); Phys.~Rev.~{\bf D55}, 2918
(1997); D.~Melikhov, hep-ph/0110087.
\bibitem{bakulev}
A.~P.~Bakulev, A.~V.~Pimikov, and N.~G.~Stefanis, Phys.\ Rev.~{\bf
D79}, 093010 (2009); Mod.~Phys.~Lett.~{\bf A24}, 2848 (2009).
\end{thebibliography}

\end{document}